\documentclass[10pt]{article}
\usepackage{graphicx}
\usepackage{amsmath}
\usepackage{amssymb}
\usepackage{graphicx}
\usepackage{dcolumn}
\usepackage{bm}
\usepackage{hyperref}
\usepackage{authblk}

\newcommand{\dslash}{\not\!\partial}

\newcommand{\vphi}{\varphi}

\def    \pt {\mbox{$p_{T} >$}}

\def    \gev {\mbox{$\mathrm{GeV}$}}

\begin{document}

\title{\vspace{-15mm}\fontsize{24pt}{10pt}\selectfont\textbf{Sensitivity to the Single Production of Vector-Like Quarks at an Upgraded Large Hadron Collider }}
\author[1]{T. Andeen }
\author[2]{C. Bernard}
\author[2]{K. Black}
\author[3]{T. Childers}
\author[2]{L. Dell'Asta}
\author[4]{N. Vignaroli}
\affil[1]{Department of Physics, Columbia  University}
\affil[2]{Department of Physics, Boston University}
\affil[3] {CERN}
\affil[4]{Department of Physics, Michigan State University}

\maketitle

\abstract{In this note we consider the sensitivity of the Large Hadron Collider (LHC) to the single production of new heavy vector-like quarks. We consider a model with large mixing with the standard model top quark with electroweak production of single heavy top quarks. We consider center of mass energies of 14, 33, and 100 TeV with various pileup scenarios and present the expected sensitivity and exclusion limits. }


\section{A benchmark model for vector-like quarks}
\label{sec:model}

Vector-like quarks (VLQs) ~\cite{Juan} are a general prediction of a wide class of beyond the Standard Model theories, as extra dimensions, composite Higgs models (see \cite{DeSimone:2012fs, Chala:2013ega, Vignaroli:2012nf, Carmona:2012jk, Vignaroli:2012sf, Bini:2011zb, Barcelo:2011wu}  for recent studies on VLQs in these frameworks), Little Higgs \cite{Han:2003wu, Perelstein:2003wd} and top-coloron models \cite{Chivukula:2013kw}.\\
We will consider as a benchmark model to describe the VLQ phenomenology a two-site description that reproduces the low-energy limit of a large set of composite Higgs models (CHM) \cite{Agashe:2004rs} and warped extra-dimensional theories with a custodial symmetry in the bulk~\cite{Agashe:2003zs}. \\
CHMs are compelling theories of new physics. They give an explanation of the EWSB by considering that it is triggered by a new strong dynamics, with a scale of compositeness of several TeVs. The Higgs is a field of the composite sector. Because of its composite nature, its mass is protected from radiative corrections above the compositeness scale and is further protected if it is also the pseudo-Goldstone boson of some symmetry breaking in the strong sector \cite{Kaplan:1983fs}. In this case the Higgs can be naturally much lighter than the other resonances from the strong sector (which have masses of the TeV order). A crucial role in this mechanism is played by the top-partner VLQs, arising from the strong sector, which intervene in cutting-off the top-loop contribution to the Higgs mass.\\  
The two building blocks of the effective theory we will consider are the weakly-coupled sector of the elementary fields and the composite sector, that comprises the Higgs. The two sectors are linearly coupled to each other through mass mixing terms \cite{Kaplan:1991dc}. After diagonalization, the elementary/composite basis rotates to the mass eigenstate one, made of SM and heavy states, among which the VLQs, that are admixture of elementary and composite modes. A minimal model, which incorporates the custodial symmetry and the Left-Right parity needed for CHM to pass the EWPT \cite{Agashe:2006at} and which includes the full set of resonances which are needed to generate the top-quark mass is that where composite fermions fill a 5 of $SO(5)$. This same description has been adopted in \cite{Contino:2008hi, Mrazek:2009yu, Vignaroli:2012nf}.  \\
The composite sector has a global symmetry $SO(4)\times U(1)_X \sim SU(2)_L\times SU(2)_R \times U(1)_X$ and includes the Higgs
\begin{equation} \label{eq:higgs}
{\cal H} = (\mathbf{2},\mathbf{2})_{0} = 
 \begin{bmatrix} \phi_0^\dagger & \phi^+ \\ - \phi^- & \phi_0 \end{bmatrix} \, ,
\end{equation}
and the following set of VLQs:

\begin{align}\label{eq:fermions} 
\begin{split}	
&	\mathcal{Q}=\left[\begin{array}{cc}
	T & T_{5/3} \\ 
	B & T_{2/3} \end{array}\right]=\left(\mathbf{2},\mathbf{2}\right)_{2/3} \ , \ \tilde{T}=\left(\mathbf{1},\mathbf{1}\right)_{2/3}
 \end{split}
\end{align}

namely, a weak singlet, $\tilde{T}$, partner of $t_R$, and two weak doublets, $(T, B)$, partner of $(t_L, b_L)$, and $(T_{5/3}, T_{2/3})$, the doublet of exotic quarks which have no direct mixing with the top. These latter, which are generally named `custodians', can be much lighter than the other VLQs in the limit case of a fully composite $t_L$.  
The elementary sector has the same particle content of the SM without the Higgs. The $SU(2)_L \times U(1)_Y$ 
elementary fields gauge the corresponding global
invariance of the composite sector, with $Y = T_{R}^3 + X$. \\
The Lagrangian that describes our model (in the gauge-less limit) reads:

\begin{align}  \label{eq:Ltotal}
{\cal L} =&  \, {\cal L}_{elementary} + {\cal L}_{composite} + {\cal L}_{mixing} \\[0.5cm]
\label{eq:Lelem}
 {\cal L}_{elementary} = &\,\bar q_L i\!\dslash\, q_L + \bar t_R i\!\dslash\, t_R \\[0.3cm]
\label{eq:Lcomp}
 {\cal L}_{composite}  = 
&\ \text{Tr}\left\{\bar{\mathcal{Q}}\left(i\dslash -M_{Q*}\right)\mathcal{Q}\right\} + \text{Tr}\left\{\bar{\tilde{T}}\left(i\dslash-M_{\tilde{T}*}\right)\tilde{T}\right\}\\
& + \frac{1}{2} \,\text{Tr}\left\{  \partial_\mu {\cal H}^\dagger \partial^\mu {\cal H} \right\} - V( {\cal H}^\dagger {\cal H}) + Y_{*}\text{Tr}\left\{ \bar{\mathcal{Q}}\mathcal{H}\right\}\tilde{T} 
  &
\\[0.3cm]
\begin{split} \label{eq:Lmixing}
{\cal L}_{mixing} =&  -\Delta_{L}\bar{q}_{L}\left(T,B\right)-\Delta_{R}\bar{t}_{R}\tilde{T}+h.c.
\end{split}
\end{align}
%
where $V( {\cal H}^\dagger {\cal H})$ is the Higgs potential. \\
$ {\cal L}^{YUK}=Y_{*}\text{Tr}\left\{ \bar{\mathcal{Q}}\mathcal{H}\right\}\tilde{T} $ describes the Yukawa interactions among Higgs and composite fermions. The Yukawa coupling $Y_{*}$ is large, $1< Y_{*} \ll 4\pi$, where $4\pi$ marks out the non-perturbative regime. $Y_*$ values greater than 3 are generally preferred by electoweak precision tests (for smaller $Y_*$ values the theory also predict lighter vector resonances that would give somehow large corrections to the S parameter \cite{Contino:2006qr}).\\
The two-site Lagrangian (\ref{eq:Ltotal}) can be diagonalized by a field rotation parametrized by:
\begin{equation}
\tan\vphi_{tR} = \frac{\Delta_{R}}{M_{\tilde{T}*}} \equiv \frac{s_{R}}{c_{R}} , \qquad
\tan\vphi_{L} =\frac{\Delta_{L}}{M_{Q*}}\equiv \frac{s_{L}}{c_{L}} 
\end{equation}
\noindent
Where $\sin\vphi_{tR}$ (shortly indicated as $s_R$) and $\sin\vphi_{L}$ ($s_L$) respectively represent the degree of compositeness of $t_R$ and $(t_L, b_L)$. 
After the diagonalization of the elementary/composite mixing, the Yukawa Lagrangian for SM and heavy fields 
reads:
 \begin{align}
\begin{split}\label{eq:Lyuk}	
\mathcal{L}^{YUK}=&+Y_{*}s_{L}c_{R}\left(\bar{t}_{L}\phi^{\dag}_{0}\tilde{T}_{R}-\bar{b}_{L}\phi^{-}\tilde{T}_{R}\right)-Y_{*}s_{R}\left(\bar{T}_{2/3L}\phi_{0}t_{R}+\bar{T}_{5/3L}\phi^{+}t_{R}\right)\\ 
& -Y_{*}c_{L}s_{R}\left(\bar{T}_{L}\phi^{\dag}_{0}t_{R}-\bar{B}_{L}\phi^{-}t_{R}\right)+Y_{*}s_{L}s_{R}\left(\bar{t}_{L}\phi^{\dag}_{0}t_{R}-\bar{b}_{L}\phi^{-}t_{R}\right)\\ &+\ h.c.+\ \dots\\
\end{split}
\end{align}
After the EWSB, the top-quark mass is generated, $m_{t}=\frac{v}{\sqrt{2}}Y_{*}s_{L}s_{R}$. The induced electroweak mixing among fermions also generates effective couplings of the VLQs with a SM quark and a weak boson. These effective couplings determine the VLQ decays and allow for the VLQ single production. \\
%
%
%
The rates for the VLQ ($\chi$) decays into a weak boson and a SM quark ($\psi$) read:

\begin{align}
\begin{split}
& \Gamma\left(\chi \rightarrow W_L\psi\right)= \frac{\lambda^2_{W\chi}}{32 \pi}M_{\chi}
\left[\left( 1+\frac{m^2_{\psi}-M^2_W}{M^2_{\chi}}\right)\left( 1+\frac{m^2_{\psi}+2M^2_W}{M^2_{\chi}}\right)-4\frac{m^2_{\psi}}{M^2_{\chi}}\right]\sqrt{1-2\frac{m^2_{\psi}+M^2_W}{M^2_{\chi}}+\frac{\left(m^2_{\psi}-M^2_W\right)^2 }{M^4_{\chi}}} \\
& \Gamma\left(\chi\rightarrow Z_L\psi\right)= \frac{\lambda^2_{Z\chi}}{64 \pi}M_{\chi}
\left[\left( 1+\frac{m^2_{\psi}-M^2_Z}{M^2_{\chi}}\right)\left( 1+\frac{m^2_{\psi}+2M^2_Z}{M^2_{\chi}}\right)-4\frac{m^2_{\psi}}{M^2_{\chi}}\right] \sqrt{1-2\frac{m^2_{\psi}+M^2_Z}{M^2_{\chi}}+\frac{\left(m^2_{\psi}-M^2_Z\right)^2 }{M^4_{\chi}}}\\
&\Gamma\left(\chi\rightarrow h\psi\right)= \frac{\lambda^2_{h\chi}}{64 \pi}M_{\chi}
\left( 1+\frac{m^2_{\psi}}{M^2_{\chi}}-\frac{M^2_{h}}{M^2_{\chi}}\right)
\sqrt{\left(1-\frac{m^2_{\psi}}{M^2_{\chi}}+\frac{M^2_{h}}{M^2_{\chi}}\right)^2-4\frac{M^4_{h}}{M^4_{\chi}}}\ .
\end{split}
\end{align}

The effectice vertices $\lambda_{W/Z/h\chi}$, that can be directly extracted from the Yukawa Lagrangian in (\ref{eq:Lyuk}), are, for the different VLQs:
\begin{center}
\begin{tabular}{c| c c c c c}\label{eq:vertices}
&  $T$\, & $B$\,  & $T_{2/3}$\; & $T_{5/3}$\; & $\tilde{T}$ \\[1mm]
\hline
 $\chi \to W \psi$ \; & $ 0$ &  $Y_* c_Ls_R$\; & $0$ \; & $Y_* s_R$\; & $Y_* s_Lc_R$ \\
 $\chi \to Z \psi$ \; & $Y_* c_Ls_R$ & $0$ & $Y_* s_R$ & $0$ & $Y_* s_Lc_R$ \\
 $\chi \to h \psi$ \; & $Y_* c_Ls_R$ & $0$ & $Y_* s_R$ & $0$ & $Y_* s_Lc_R$
\end{tabular} 
\end{center}

%

 The above expressions are calculated by diagonalizing the mixing among fermions at first order in   
\begin{equation}
 \sin\theta_{\chi}=\frac{\lambda_{\chi} v}{\sqrt{2}M_{\chi}} \ ,
\end{equation}
which parametrizes the superposition of a VLQ $\chi$ with the top. We expect corrections of $\mathcal{O}(1)$ in the VLQ decay rates and in the VLQ single production cross sections, for $\lambda_{\chi} v/(\sqrt{2}M_{\chi})\simeq 1$. \footnote{To extract the exact value of $\sin\theta$, one should fully diagonalize the 4x4 matrix of the mixing of all the 2/3 charge fermions. In the case of a fully composite $t_L$ and for $\lambda_{\tilde{T}}$=3 and $M_{\tilde{T}}=500$ GeV, corresponding to a value $\sin\theta^{LO}_{\tilde{T}}=\lambda_{\tilde{T}} v/(\sqrt{2}M_{\tilde{T}})=1$, we find, for example, a correction $(\sin\theta^{LO}_{\tilde{T}}/\sin\theta^{FULL}_{\tilde{T}})^2\simeq 1.7$. }

Since VLQs are essentially composite states which couple strongly to composite modes and thus to longitudinally polarized weak bosons and to the Higgs, their branching-ratios are basically fixed by the equivalence theorem. One finds, approximately

\begin{center}
\begin{tabular}{c| c c c c c}\label{eq:vertices}
&  $T$\, & $B$\,  & $T_{2/3}$\; & $T_{5/3}$\; & $\tilde{T}$ \\[1mm]
\hline 
 $BR[\chi \to W \psi]$ \; & $ 0$ &  $1$\; & $0$ \; & $1$\; & $0.5$ \\
 $BR[\chi \to Z \psi]$ \; & $0.5$ & $0$ & $0.5$ & $0$ & $0.25$ \\
 $BR[\chi \to h \psi]$ \; & $0.5$ & $0$ & $0.5$ & $0$ & $0.25$
\end{tabular} 
\end{center}

\begin{figure}[]
\begin{center}
\includegraphics[width=0.49\textwidth,clip,angle=0]{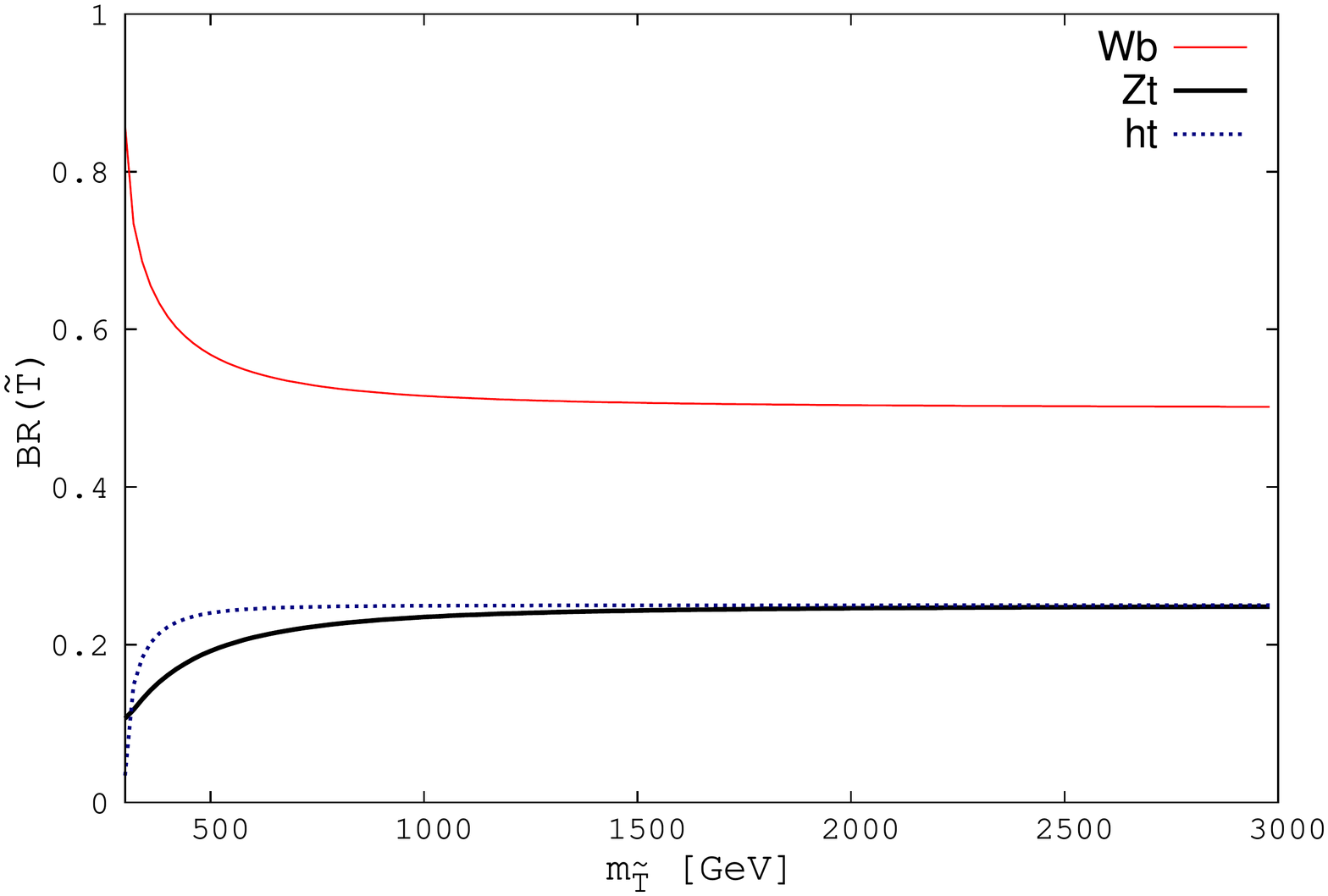}
\includegraphics[width=0.49\textwidth,clip,angle=0]{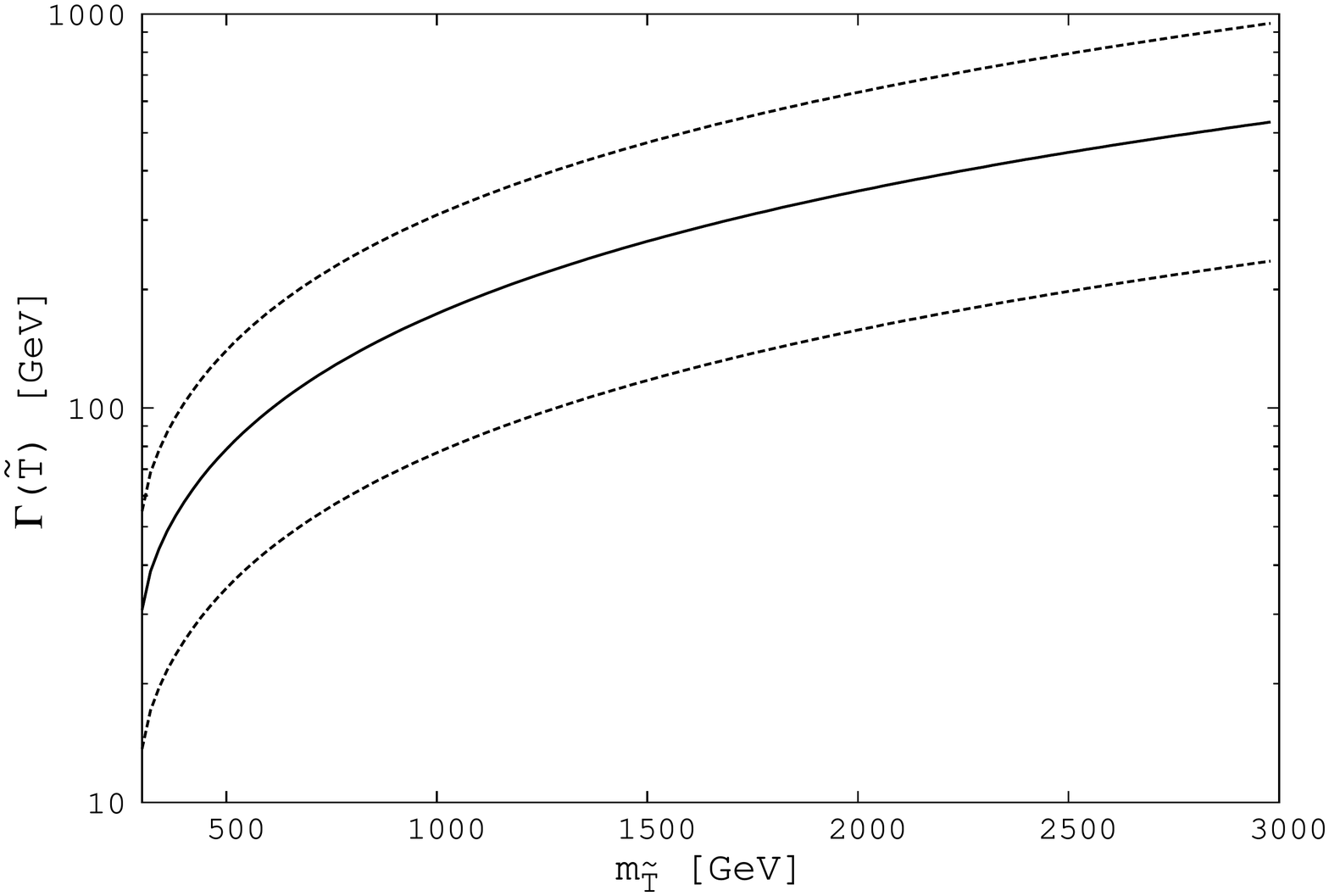}
\caption[]{
\label{fig:Ts-decay}
\small
BRs (Left Plot) and total decay width (Right Plot) of $\tilde{T}$. 
The width depends quadratically on $\lambda_{\tilde{T}}$; 
the continuous line in the Right Plot refers to $\lambda_{\tilde{T}}=3$, the dotted lines define a range of variation $2<\lambda_{\tilde{T}}<4$ of the total decay width. 
}
\end{center}
\end{figure}

Fig. \ref{fig:Ts-decay} shows the decay branching-ratios and the total decay width for $\tilde{T}$. VLQ total decay widths depend quadratically on the effective vertices $\lambda_{\chi}$.

The flavor structure of the two-site model here described has been analyzed in \cite{Vignaroli:2012si, Barbieri:2012tu, Straub:2013zca, Redi:2011zi}. In the anarchic scenario for the flavor of the composite sector, where $Y_*$ is a matrix in the flavor space with elements all of the same size, flavor observables like $\epsilon_K$ and $\epsilon^{'}/\epsilon_K$ place strong constraints on the CHM spectrum. Recent studies \cite{Barbieri:2012tu, Redi:2012uj} have shown that the flavor constraints on VLQ masses can be lowered to the order of 1 TeV or below if a $U(2)^3$ flavor symmetry is present in the composite sector, instead of the anarchic flavor structure. In this case, VLQs couple strongly to third-generation quarks and weakly to light quarks, thus reflecting the same VLQ phenomenology above described.

\section{VLQ production mechanisms}

\begin{figure}[]
\begin{center}
\includegraphics[width=0.5\textwidth,clip,angle=0]{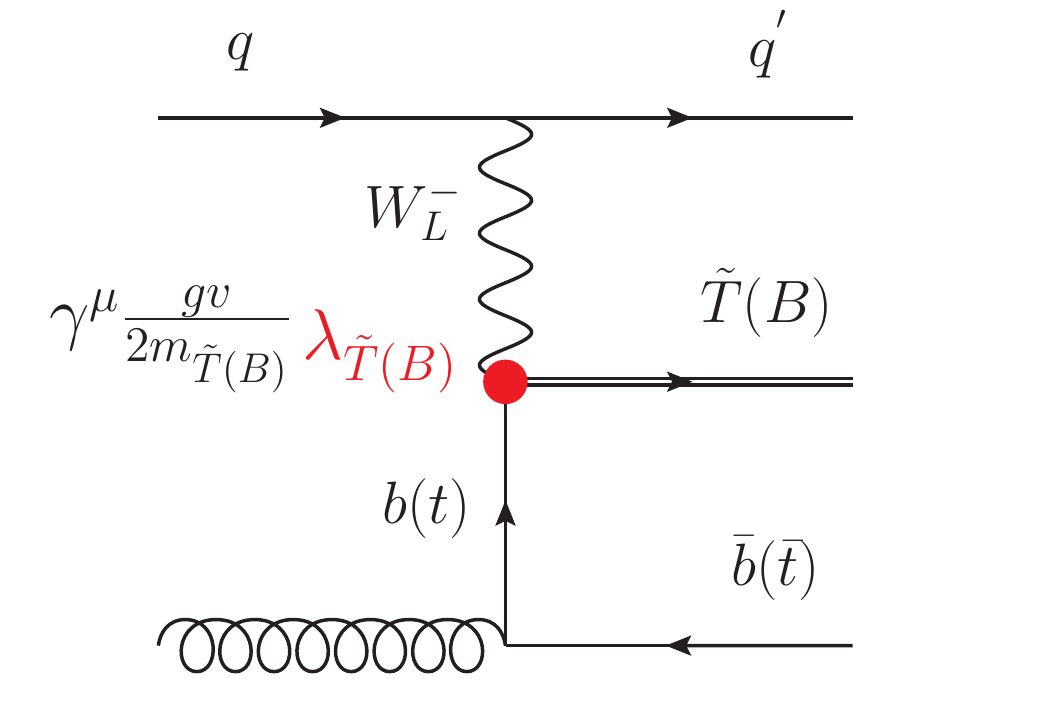}
\caption[]{
\label{fig:single-prod}
\small
Single production of top- and bottom- prime VLQs. 
}
\end{center}
\end{figure}

 \begin{figure}[]
\begin{center}
\includegraphics[width=0.6\textwidth,clip,angle=0]{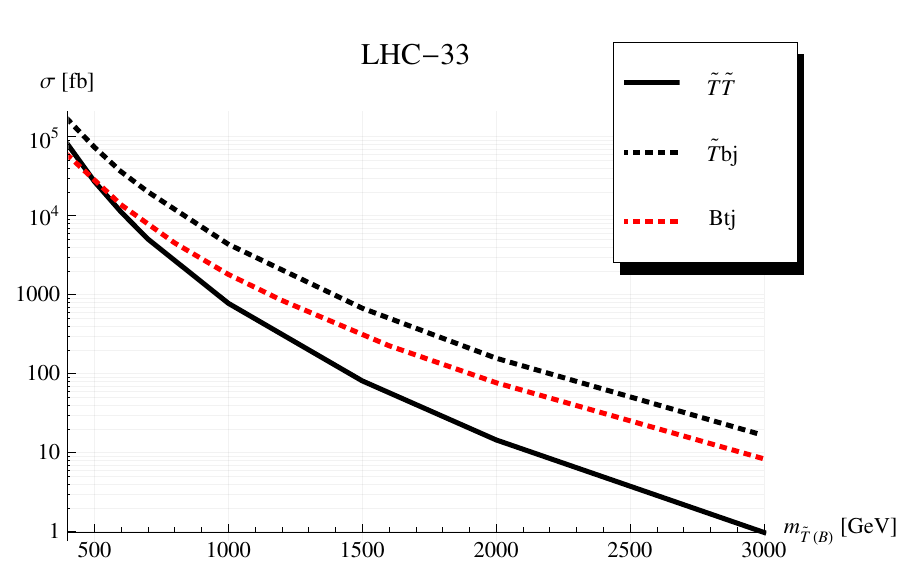}
\caption[]{
\label{fig:xsec}
\small
VLQ cross sections at the 33 TeV LHC. Dotted curves correspond to single-productions for $\lambda_{\chi}=3$.
}
\end{center}
\end{figure}
  
VLQs can be produced at the LHC in pairs, the dominant mode is via gluon-gluon fusion, or singly, by means of their electroweak effective couplings to a weak boson and a SM quark (Fig. \ref{fig:single-prod}). These latter production mechanisms have larger rates than those of pair productions for heavier VLQs. Moreover, analyses of single-production channels might permit the measurement of the effective couplings $\lambda_{\chi}$. Fig. \ref{fig:xsec} shows the cross sections at the 33 TeV LHC for VLQ pair production and for single production of $\tilde{T}$ and $B$ (the same for $T_{5/3}$), for $\lambda_{\chi}=3$. The $\tilde{T}$ single-production, that proceeds via the intermediate exchange of a bottom quark \footnote{$\tilde{T}$ single-production can also occur at leading-order in QCD couplings, from an initial b-quark parton. In this case, however, there would not be the `extra' b-quark in the final state, which is instead very useful to handle the SM background. },  has a rate significantly higher than those of $B$ and $T_{5/3}$ single productions, which are mediated by the exchange of a top.

\section{Simulation}

In order to evaluate the sensitivity , samples were generated using the DELPHES \cite{delphes} fast detector simulation using the generic ``Snowmass Detector" parameters \cite{snowdet}.
The background samples were generated in bins of $H_{T}$, as described in \cite{samples}. The signal samples were produced using the model from one of the authors in Ref \cite{Vignaroli:2012nf} as described
in the Section \ref{sec:model}. 

The following scenarios were considered:
\begin{itemize}
\item Three scenarios at $\sqrt{s} $= 14 TeV, with an average of 0, 50, and 140 pileup events with an integrated luminosity of 3000 $\rm fb^{-1}$.
\item Three scenarios at $\sqrt{s}$ = 30 TeV, with an average of 0, 50, and 140 pileup events with an integrated luminosity of 3000  $ \rm fb^{-1}$.
\item Three scenarios at $\sqrt{s}$ = 100 TeV, with an average of 0, 50, and 140 pileup events with an integrated luminosity of 3000 $ \rm fb^{-1}$.
\end{itemize} 

\section{Channels Considered} 

As mentioned in Section \ref{sec:model} , the heavy like top quark can decay into one of three modes:
\begin{itemize}
\item{ T $\rightarrow$ Wb }
\item{T $\rightarrow$ tZ}
\item{T $\rightarrow$ tH}
\end{itemize}

In this note we consider the later two decay modes.

\subsection{ T $\rightarrow$ tH} 

In this decay mode we focus on the decay of the Higgs to its most frequent decay into $b \bar{b}$ and the case where the top quark decays semi-leptonically into a charged lepton, neutrino and b-quark.
We reconstruct the heavy VLQ in a series of steps. First we select events based on the decay topology:

\begin{itemize}
\item Events are required to have one lepton (muon or electron) with $|\eta |<$ 2.5
\item Events are required to have missing transverse energy greater than 30 GeV
\item Events are required to have at three jets that have $p_{T} >$ 25 GeV that are identified as b-jets 
\item Events are required to have two jets with $|\eta| >$ 3.0 (from the forward scattered quarks from the hard subprocess)
\item Events are required to have $H_{T}> $ 750 GeV
\end{itemize}

Events are then reconstructed in the following manner. The charged lepton and the neutrino are assumed to come from the W boson in the decay. We measure the complete four-vector of the
charged lepton but only can assume the two transverse components of the missing transverse energy are from the two transverse components of the neutrino. Since we are missing the longitudinal component
of the neutrino momentum we must make some assumption to completely reconstruct the event. We force the invariant mass of the missing transverse momentum and the charged lepton to that of the pole mass of the W. This allows for us to solve for the longitudinal component of the neutrino momentum. This is a quadratic equation and leads to up to two real solutions. In the case of two solutions we take the solution which 
minimizes the angle between the charged lepton and the neutrino. 

After reconstruction of the top quark candidate in the event we make the following additional requirements:

\begin{itemize}
\item The $\Delta$ R between the lepton and the b-jet used for the top reconstruction must be larger than 0.7
\item One Cambridge-Achen jet must have an invariant mass between 100-150 GeV
\end{itemize}

We utilize the entire Snowmass background samples to estimate the background but by far the largest contribution is standard model $t \bar{t}$ production. 

\section{T $\rightarrow$tZ}

For the tZ channel we focus on the signal with the lowest background and select trilepton events. In this decay mode we focus on the decay of the $Z$ boson into leptons and the case where the top quark decays semi-leptonically into a charged lepton, neutrino and b-quark. 

The reconstructed objects used for the analysis are selected as following:
\begin{itemize}
\item Jets are reconstructed using the anti$-k_t$ algorithm with $r = 0.5$. They are required to have \pt higher than 30 \gev and $|\eta |<$ 5.0. Jets must be isolated, therefore an overlap removal between jets is applied, in $\Delta R < 0.5$: the jet with the higest \pt is retained and the other discarded.
\item Leptons (muons or electrons) are required to have \pt higher than 20 \gev and $|\eta |<$ 2.5. Furthermore they are required to be isolated from jets, within $\Delta R < 0.5$.
\end{itemize}
A b-tagging algorithm is also available to identify jets coming from a b-quark. 

We reconstruct the heavy VLQ in a series of steps. First we select events based on the decay topology:

\begin{itemize}
\item Events are required to have exactly three leptons
\item Events are required to have at least one light jet (not b-tagged)
\item Events are required to have at least two b-tagged jets
\item Events are required to have missing transverse energy greater than 30 \gev
\end{itemize}

The event reconstruction in this case is a bit simpler than the $Ht$ case. 

First of all, the $Z$ boson candidate is reconstructed from a pair of same flavor leptons and a cut on the invariant mass of the two leptons is applied. If in the events there are only two leptons with the same flavor (two electrons and a muon or viceversa), the two same flavor leptons are required to form an invariant mass within a 10 \gev window of the mass of the $Z$ boson. If instead the three leptons have all the same flavor, the couple with the invariant mass closer to the $Z$ boson mass are considered for the 10 \gev window mass cut.

After reconstructing the $Z$ boson, the light jet from the forward scattered quarks from the hard subproces is looked for. The light jet with the highest $\eta$ is selected and this jet is required to have $|\eta| > 2.5$.

The top candidate is reconstructed using the same method as described in the previous section, by using the third lepton in the event and the b-tagged jet that best reproduces the top mass. Furthermore a cut on the mass of the $Wj$ is applied: $160 \gev < m_{Wj} < 190 \gev$.

Again we utilize the entire Snowmass background samples to estimate the background. The samples with the largest contributions come from diboson production and top quark pair production associated with a boson production.

\section{Cross-Sections and Event Yields}

In Table \ref{tab:sig} and \ref{tab:xsecbkg} the cross sections for the signal and the main backgrounds respectively are reported, for the three center of mass energies considered in this study.

\section{Results}

\begin{figure}
\includegraphics[width=0.6\textwidth,clip,angle=0]{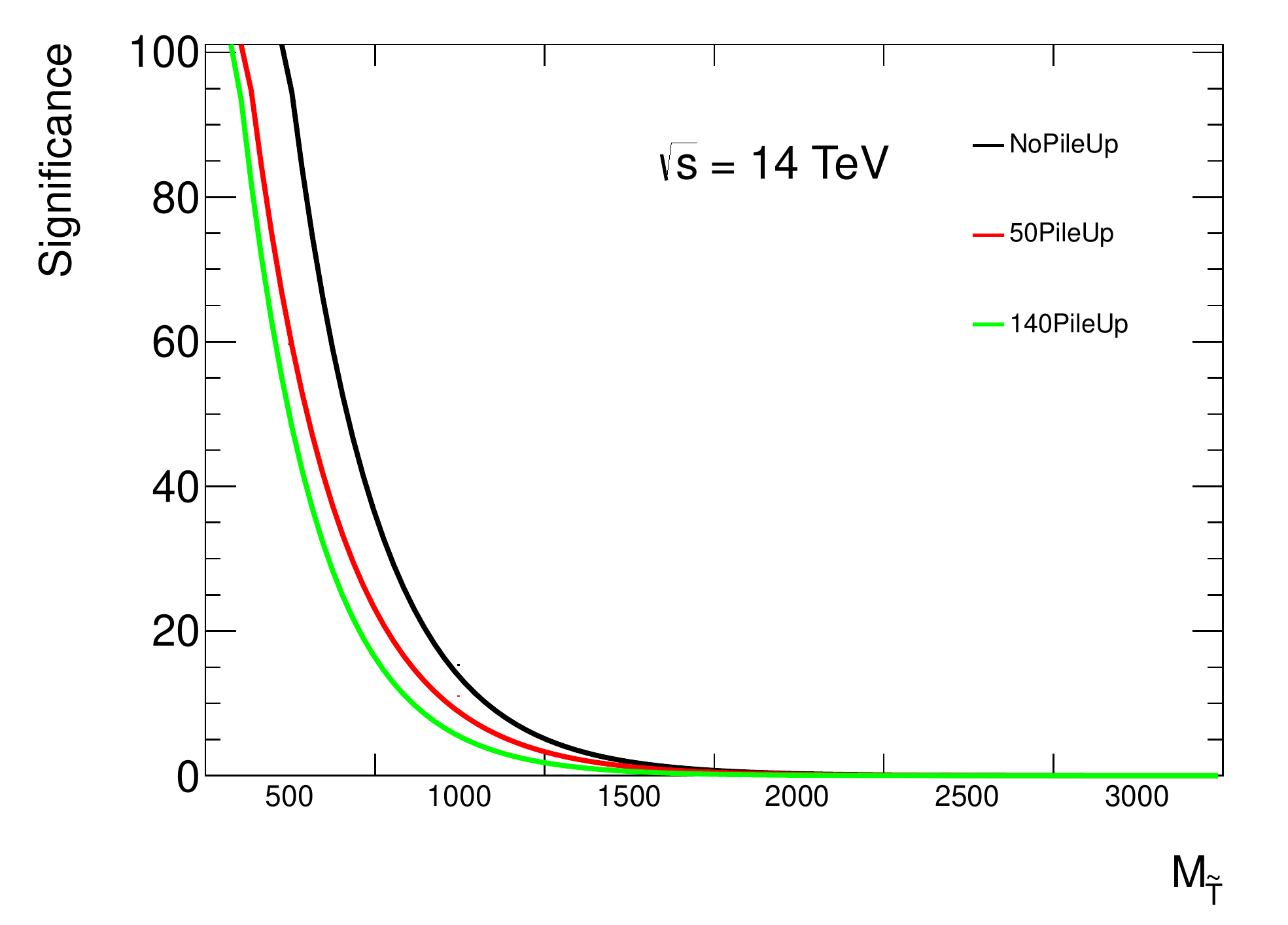}
\includegraphics[width=0.6\textwidth,clip,angle=0]{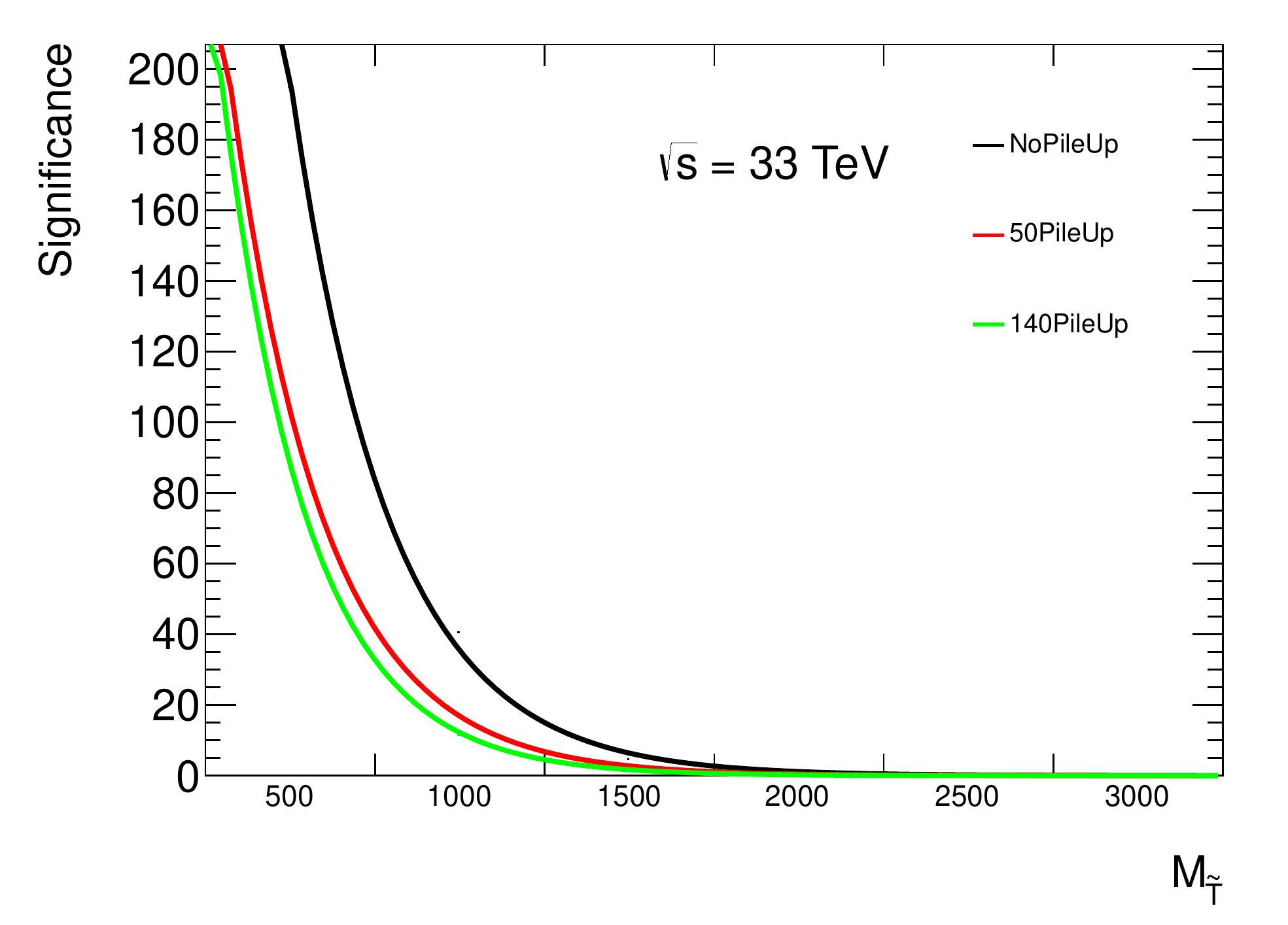}
\includegraphics[width=0.6\textwidth,clip,angle=0]{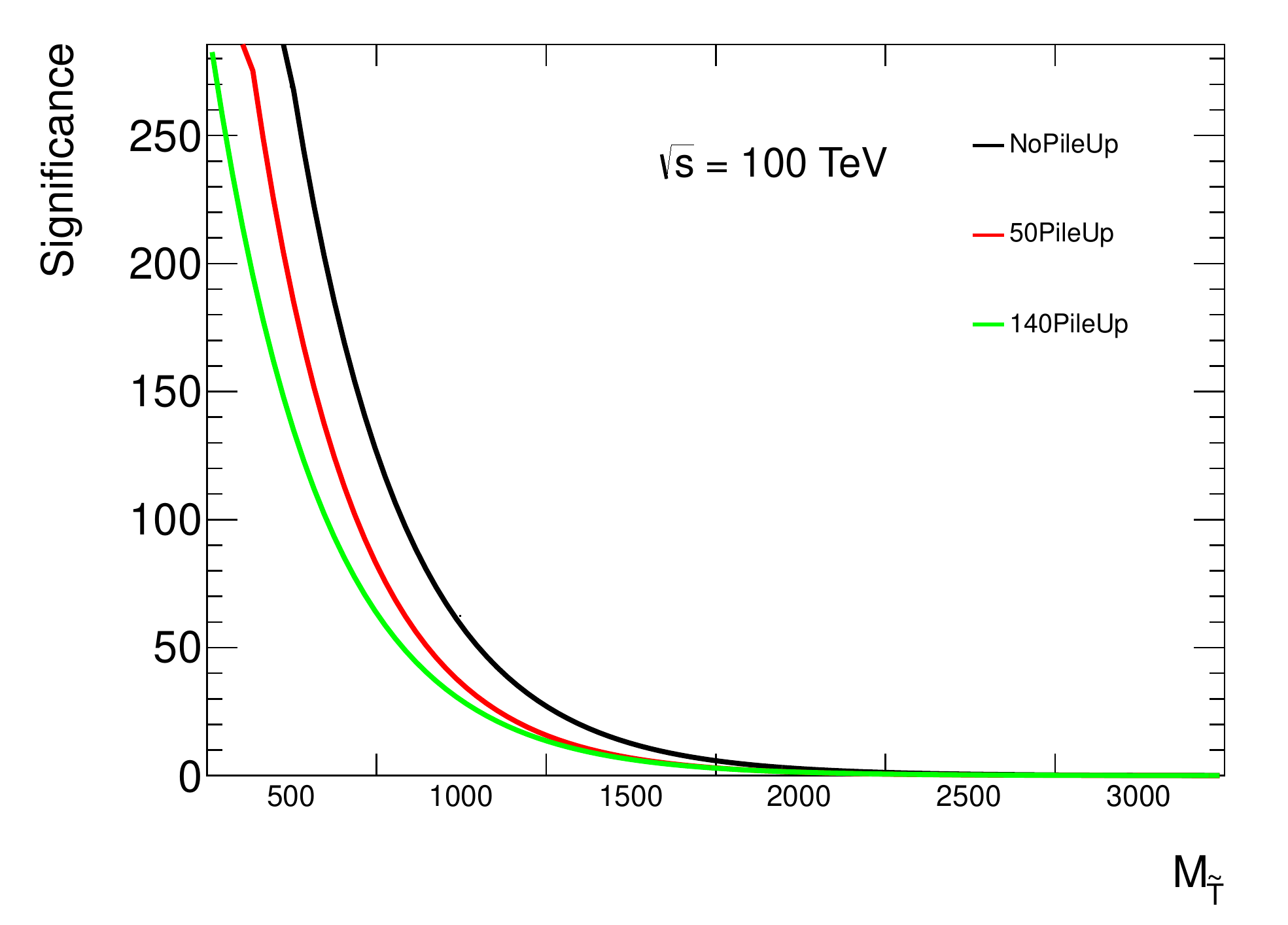}
\caption{Expected Significance at center of mass energies 14 (top left), 33 (top right), and 100 (bottom) GeV with no pileup, 50 interaction, and 140 interactions ~\label{fig:sig}.}
\end{figure}

The expected significances for the combination of both channels is shown in Figure ~\ref{fig:sig}.

\appendix

\newpage
\section{Signal Cross Sections}

\begin{table}[h!]
  \begin{center}
    \begin{tabular}{c c c c c c}
      \hline 
      \hline
      Center of & Lambda  &  Decay  &  Mass & Width & Cross\\
      Mass Energy [TeV] & & & [GeV] & [GeV] & Section [pb] \\
      \hline	
      \hline	
      100  &   3   &  Zt  &   500   &   78.6   &    117.104966 \\
      100  &   3   &  Zt  &   1000   &  173.6   &    10.877649 \\
      100  &   3   &  Zt  &   1500   &  264.9   &    2.237840 \\
      100  &   3   &  Zt  &   2000   &  355.4   &    0.675062 \\
      100  &   3   &  Zt  &   2500   &  445.4   &    0.254231 \\
      100  &   3   &  Zt  &   3000   &  535.3   &    0.110822 \\
      \hline	
      33  &   3   &  Zt  &   500   &   78.6   &    17.296388 \\
      33  &   3   &  Zt  &   1000   &   78.6   &    2.113275 \\
      33  &   3   &  Zt  &   1500   &  173.6   &    0.257407 \\
      33  &   3   &  Zt  &   2000   &  264.9   &    0.057440 \\
      33  &   3   &  Zt  &   2500   &  355.4   &    0.016986 \\
      33  &   3   &  Zt  &   3000   &  535.3   &    0.005486 \\
      \hline	
      14  &   3   &  Zt  &   500   &   78.6   &    2.573795 \\
      14  &   3   &  Zt  &   1000   &  173.6   &    0.111849 \\
      14  &   3   &  Zt  &   1500   &  264.9   &    0.011920 \\
      14  &   3   &  Zt  &   2000   &  355.4   &    0.001939 \\
      14  &   3   &  Zt  &   2500   &  445.4   &    0.000402 \\
      14  &   3   &  Zt  &   3000   &  535.3   &    0.000098 \\
      \hline 
      \hline
    \end{tabular}
    \caption{Generated Masses, Widths and Cross Sections for top partner single production decaying into $Zt$.}
    \label{tab:xsecsigZt}
  \end{center}
\end{table}

\begin{table}
  \begin{center}
    \begin{tabular}{c c c c c c}
      \hline 
      \hline
      Center of & Lambda  &  Decay  &  Mass & Width & Cross\\
      Mass Energy [TeV] & & & [GeV] & [GeV] & Section [pb] \\
      \hline	
      \hline	
      100  &   3   &  Wb  &   500   &   78.6   &    318.691708 \\
      100  &   3   &  Wb  &   1000   &  173.6   &    23.474201 \\
      100  &   3   &  Wb  &   1500   &  264.9   &    4.624145 \\
      100  &   3   &  Wb  &   2000   &  355.4   &    1.371373 \\
      100  &   3   &  Wb  &   2500   &  445.4   &    0.513678 \\
      100  &   3   &  Wb  &   3000   &  535.3   &    0.223635 \\
      \hline	
      33  &   3   &  Wb  &   500   &   78.6   &    47.820304 \\
      33  &   3   &  Wb  &   1000   &   78.6   &    5.122861 \\
      33  &   3   &  Wb  &   1500   &  173.6   &    0.540823 \\
      33  &   3   &  Wb  &   2000   &  264.9   &    0.117806 \\
      33  &   3   &  Wb  &   2500   &  355.4   &    0.034511 \\
      33  &   3   &  Wb  &   3000   &  535.3   &    0.011076 \\
      \hline	
      14  &   3   &  Wb  &   500   &   78.6   &    7.276922 \\
      14  &   3   &  Wb  &   1000   &  173.6   &    0.244653 \\
      14  &   3   &  Wb  &   1500   &  264.9   &    0.024868 \\
      14  &   3   &  Wb  &   2000   &  355.4   &    0.003980 \\
      14  &   3   &  Wb  &   2500   &  445.4   &    0.000819 \\
      14  &   3   &  Wb  &   3000   &  535.3   &    0.000198 \\
      \hline 
      \hline
    \end{tabular}
    \caption{Generated Masses, Widths and Cross Sections for top partner single production decaying into $Wb$.}
    \label{tab:xsecsigWb}
  \end{center}
\end{table}

\begin{table}
  \begin{center}
    \begin{tabular}{c c c c c c}
      \hline 
      \hline
      Center of & Lambda  &  Decay  &  Mass & Width & Cross\\
      Mass Energy [TeV] & & & [GeV] & [GeV] & Section [pb] \\
      \hline	
      \hline	
      100  &   3   &  ht  &   500   &   78.6   &    107.749943 \\
      100  &   3   &  ht  &   1000   &  173.6   &    10.286034 \\
      100  &   3   &  ht  &   1500   &  264.9   &    2.256206 \\
      100  &   3   &  ht  &   2000   &  355.4   &    0.726233 \\
      100  &   3   &  ht  &   2500   &  445.4   &    0.290587 \\
      100  &   3   &  ht  &   3000   &  535.3   &    0.134324 \\
      \hline	
      33  &   3   &  ht  &   500   &   78.6   &    17.063678 \\
      33  &   3   &  ht  &   1000   &   78.6   &    2.175484 \\
      33  &   3   &  ht  &   1500   &  173.6   &    0.296841 \\
      33  &   3   &  ht  &   2000   &  264.9   &    0.074993 \\
      33  &   3   &  ht  &   2500   &  355.4   &    0.025288 \\
      33  &   3   &  ht  &   3000   &  535.3   &    0.009543 \\
      \hline	
      14  &   3   &  ht  &   500   &   78.6   &    2.780178 \\
      14  &   3   &  ht  &   1000   &  173.6   &    0.140249 \\
      14  &   3   &  ht  &   1500   &  264.9   &    0.018871 \\
      14  &   3   &  ht  &   2000   &  355.4   &    0.004187 \\
      14  &   3   &  ht  &   2500   &  445.4   &    0.001291 \\
      14  &   3   &  ht  &   3000   &  535.3   &    0.000507 \\
      \hline 
      \hline
    \end{tabular}
    \caption{Generated Masses, Widths and Cross Sections for top partner single production decaying into $Ht$.}
    \label{tab:xsecsigHt}
  \end{center}
\end{table}

\newpage
\section{Background Cross Sections}

\begin{table}[h!]
  \begin{center}
    \begin{tabular}{c c c c}
      \hline 
      \hline
      Center of  &  Process  &  $H_T$ bin & Cross\\
      Mass Energy [TeV] & & [GeV] & Section [pb] \\
      \hline	
      \hline	
      100 & tt & 0-1000 &    29141.30000\\
      100 & tt & 1000-2000 &     1777.28000\\
      100 & tt & 2000-3500 &      185.21600\\
      100 & tt & 3500-5500 &       18.91940\\
      100 & tt & 5500-8500 &        2.38751\\
      100 & tt & 8500-100000 &        0.27715\\
      \hline 
      33 & tt & 0-600 &   3438.70635\\
      33 & tt & 600-1200 &    505.82210\\
      33 & tt & 1200-2000 &     61.81892\\
      33 & tt & 2000-3200 &      7.65752\\
      33 & tt & 3200-4800 &      0.72643\\
      33 & tt & 4800-100000 &      0.07147\\
      \hline	
      14 & tt & 0-600 &    530.89358 $\pm$    0.15615\\
      14 & tt & 600-1100 &     42.55351 $\pm$    0.01367\\
      14 & tt & 1100-1700 &      4.48209 $\pm$    0.00164\\
      14 & tt & 1700-2500 &      0.52795 $\pm$    0.00019\\
      14 & tt & 2500-100000 &      0.05449 $\pm$    0.00002\\
      \hline 
      \hline
    \end{tabular}
    \caption{}
    \label{tab:xsecbkg}
  \end{center}
\end{table}

\begin{table}[h!]
  \begin{center}
    \begin{tabular}{c c c c}
      \hline 
      \hline
      Center of  &  Process  &  $H_T$ bin & Cross\\
      Mass Energy [TeV] & & [GeV] & Section [pb] \\
      \hline	
      \hline	
      33 & Bjj & 0-800 &    302.55913\\
      33 & Bjj & 1600-3000 &      1.73825\\
      33 & Bjj & 3000-4800 &      0.13606\\
      33 & Bjj & 4800-100000 &      0.01623\\
      33 & Bjj & 800-1600 &     16.41152\\
      \hline 
      14 & Bjj & 0-700 &     86.45604 $\pm$    0.02382\\
      14 & Bjj & 700-1400 &      4.34869 $\pm$    0.00166\\
      14 & Bjj & 1400-2300 &      0.32465 $\pm$    0.00015\\
      14 & Bjj & 2300-3400 &      0.03032 $\pm$    0.00004\\
      14 & Bjj & 3400-100000 &      0.00313 $\pm$    0.00001\\
      \hline 
      \hline
    \end{tabular}
    \caption{}
    \label{tab:xsecbkg}
  \end{center}
\end{table}

\begin{table}[h!]
  \begin{center}
    \begin{tabular}{c c c c}
      \hline 
      \hline
      Center of  &  Process  &  $H_T$ bin & Cross\\
      Mass Energy [TeV] & & [GeV] & Section [pb] \\
      \hline	
      \hline	
      100 & tB & 0-1000 &     3399.65000\\
      100 & tB & 1000-2000 &      165.25400\\
      100 & tB & 2000-3500 &       15.57060\\
      100 & tB & 3500-6000 &        1.58664\\
      100 & tB & 6000-9000 &        0.10670\\
      100 & tB & 9000-100000 &        0.01283\\
      \hline 
      33 & tB & 0-600 &    432.35695\\
      33 & tB & 600-1200 &     53.97997\\
      33 & tB & 1200-2000 &      5.60692\\
      33 & tB & 2000-3200 &      0.62643\\
      33 & tB & 3200-100000 &      0.05937\\
      \hline	
      14 & tB & 0-500 &     63.88923 $\pm$    0.01952\\
      14 & tB & 500-900 &      7.12172 $\pm$    0.00278\\
      14 & tB & 900-1500 &      0.98030 $\pm$    0.00034\\
      14 & tB & 1500-2200 &      0.08391 $\pm$    0.00004\\
      14 & tB & 2200-100000 &      0.00953 $\pm$    0.00000\\
      \hline 
      \hline
    \end{tabular}
    \caption{}
    \label{tab:xsecbkg}
  \end{center}
\end{table}

\begin{table}[h!]
  \begin{center}
    \begin{tabular}{c c c c}
      \hline 
      \hline
      Center of  &  Process  &  $H_T$ bin & Cross\\
      Mass Energy [TeV] & & [GeV] & Section [pb] \\
      \hline	
      \hline	
      100 & ttB & 0-1500 &      206.00600\\
      100 & ttB & 1500-3000 &       12.57990\\
      100 & ttB & 3000-5500 &        1.18355\\
      100 & ttB & 5500-9000 &        0.09190\\
      100 & ttB & 9000-100000 &        0.00908\\
      \hline 
      33 & ttB & 0-1200 &     23.91162\\
      33 & ttB & 1200-2200 &      1.57640\\
      33 & ttB & 2200-3600 &      0.16155\\
      33 & ttB & 3600-100000 &      0.01684\\
      \hline	
      14 & ttB & 0-900 &      2.66730 $\pm$    0.0\\
      14 & ttB & 900-1600 &      0.25047 $\pm$    0.0\\
      14 & ttB & 1600-2500 &      0.02374 $\pm$    0.0 \\
      14 & ttB & 2500-100000 &      0.00209 $\pm$    0.0\\
      \hline 
      \hline
    \end{tabular}
    \caption{}
    \label{tab:xsecbkg}
  \end{center}
\end{table}

\begin{table}[h!]
  \begin{center}
    \begin{tabular}{c c c c}
      \hline 
      \hline
      Center of  &  Process  &  $H_T$ bin & Cross\\
      Mass Energy [TeV] & & [GeV] & Section [pb] \\
      \hline	
      \hline	
      100 & BB & 0-500 &     2867.87000\\
      100 & BB & 500-1500 &      405.20000\\
      100 & BB & 1500-3000 &       22.84390\\
      100 & BB & 3000-5500 &        2.22112\\
      100 & BB & 5500-9000 &        0.20005\\
      100 & BB & 9000-100000 &        0.02441\\
      \hline 
      33 & BB & 0-400 &    776.00399\\
      33 & BB & 400-1000 &    106.85023\\
      33 & BB & 1000-2000 &     10.16835\\
      33 & BB & 2000-3400 &      0.86136\\
      33 & BB & 3400-100000 &      0.09507\\
      \hline	
      14 & BB & 0-300 &    249.97710 $\pm$    0.05919\\
      14 & BB & 300-700 &     35.23062 $\pm$    0.01132\\
      14 & BB & 700-1300 &      4.13743 $\pm$    0.00150\\
      14 & BB & 1300-2100 &      0.41702 $\pm$    0.00019\\
      14 & BB & 2100-100000 &      0.04770 $\pm$    0.00005\\
      \hline 
      \hline
    \end{tabular}
    \caption{}
    \label{tab:xsecbkg}
  \end{center}
\end{table}

\begin{table}[h!]
  \begin{center}
    \begin{tabular}{c c c c}
      \hline 
      \hline
      Center of  &  Process  &  $H_T$ bin & Cross\\
      Mass Energy [TeV] & & [GeV] & Section [pb] \\
      \hline	
      \hline	
      100 & BBB & 0-1000 &       34.45440\\
      100 & BBB & 1000-3000 &        1.86207\\
      100 & BBB & 3000-6000 &        0.08519\\
      100 & BBB & 6000-100000 &        0.00726\\
      \hline 
      33 & BBB & 0-800 &      8.68026\\
      33 & BBB & 800-2000 &      0.44927\\
      33 & BBB & 2000-3600 &      0.02699\\
      33 & BBB & 3600-100000 &      0.00246\\
      \hline	
      14 & BBB & 0-600 &      2.57304 $\pm$    0.00071\\
      14 & BBB & 600-1300 &      0.14935 $\pm$    0.00005\\
      14 & BBB & 1300-100000 &      0.01274 $\pm$    0.00001\\
      \hline 
      \hline
    \end{tabular}
    \caption{}
    \label{tab:xsecbkg}
  \end{center}
\end{table}

\end{document}